\documentclass[aps,pra,superscriptaddress,twocolumn,showpacs]{revtex4}
\usepackage{amsmath}
\usepackage{graphicx}
\usepackage{amssymb}
\usepackage{bbm, color}

\newcommand*{\eps}{\varepsilon}
\newcommand*{\cl}[1]{{\mathcal{#1}}}
\newcommand*{\bb}[1]{{\mathbb{#1}}}
\newcommand{\ket}[1]{\left|#1\right>}
\newcommand{\bra}[1]{\left<#1\right|}

\newcommand{\proj}[2]{| #1 \rangle\!\langle #2 |}
\newcommand*{\tn}[1]{{\textnormal{#1}}}
\newcommand*{\1}{{\mathbbm{1}}}
\newcommand{\T}{\mbox{$\textnormal{Tr}$}}

\begin{document}

\title{Purification of Gaussian maximally mixed states}

\author{Kabgyun Jeong}
\affiliation{Center for Macroscopic Quantum Control, Department of Physics and Astronomy, \\
Seoul National University, Seoul 08826, Korea}
\affiliation{School of Computational Sciences, Korea Institute for Advanced Study, \\
Seoul 02455, Korea}
\author{Youngrong Lim}
\affiliation{Center for Macroscopic Quantum Control, Department of Physics and Astronomy, \\
Seoul National University, Seoul 08826, Korea}

\date{\today}
\pacs{03.67.-a, 03.65.Ud, 03.67.Bg, 42.50.-p}

\begin{abstract}
We find that the purifications of several Gaussian maximally mixed states (GMMSs) correspond to some Gaussian maximally entangled states (GMESs) in the continuous-variable regime. Here, we consider a two-mode squeezed vacuum (TMSV) state as a purification of the thermal state and construct a general formalism of the Gaussian purification process. Moreover, we introduce other kind of GMESs via the process. All of our purified states of the GMMSs exhibit Gaussian profiles; thus, the states show maximal quantum entanglement in the Gaussian regime.
\end{abstract}

\maketitle

\section{Introduction} \label{intro}
The principle of quantum purification means that for any mixed quantum state of a system $A$ with a given orthonormal basis, there \emph{exists} an orthonormal basis for ancillary system $B$ (with at least the same dimension as that of the system $A$) that corresponds to the orthonormal basis of the system $A$. These two bases are closely related by a local unitary operation on the ancillary system $B$. This statement is the famous Hughston-Jozsa-Wootters (HJW) theorem~\cite{HJW93}. For example, purification of a $d$-dimensional maximally mixed state (MMS) is just one of the $d\times d$-dimensional maximally entangled states (MESs) up to the local unitary operations on the ancillary system. In quantum information theory, purification is a mathematical procedure for generating a pure state from a mixed state~\cite{J94,P98,NC00}; however, this concept differs from the purification in which a pure state is constructed from a mixed state via many copies of mixed states with the same dimensions~\cite{BDSW96,BBPSSW97,Yubo13}. For a continuous-variable (CV) system, the concept of MMS is rather vague and still not well-understood. Instead of exhaustively considering all possible CV systems, here we focus on the Gaussian CV systems that have many practical applications in quantum optics and quantum information fields~\cite{CV_QT,Loock07}. Note that the ``Gaussian state'' here means a quantum state having a Gaussian profile in the phase space, i.e., its Wigner function is a Gaussian distribution. We investigate several Gaussian maximally mixed states (GMMSs) and their purified states, i.e., the Gaussian maximally entangled states (GMESs). We call this process and its underlying principle  $g$-\emph{purification} and \emph{Gaussian MMS-MES correspondence}, respectively.

It is noteworthy that any Gaussian state can be decomposed by an infinite-dimensional Fock basis, and any convex combination of quantum states gives a quantum state again. By using the correspondence, we find a \emph{new} class of GMESs such as the (known) two-mode squeezed vacuum (TMSV) state with infinite squeezing parameter which is the purification of the thermal state with infinite temperature, as well as the $g$-purified MESs over Br\'{a}dler's CV MMSs~\cite{B05} and squeezed MMSs~\cite{JKL15} in the Gaussian regime. These are then generalized in a single statement (see below). Furthermore, this method can be a powerful tool for Gaussian quantum information~\cite{WPGCRSL12,ARL14} (and references therein).

While the amount of entanglement of a given Gaussian state with a given purity (or mixedness) can be calculated~\cite{Adesso2004}, a GMMS in the CV regime that gives the MES via the purification process is not precisely defined. Therefore, we suggest several GMMS candidates (depicted in FIG.~\ref{Figure-1} below) and investigate their $g$-purifications explicitly. Note that the exact MMS is present only in a bounded Hilbert space. Even if we are dealing with an unbounded Hilbert space, however, we first perform the calculations in the bounded region and then take a limit of that region to infinity. Moreover, we describe an equivalence relation for GMMSs in the limit of the spectrum of the number operator $\hat{n}$ tending to infinity. Prior to the study of Gaussian MMSs or MESs, we briefly review the MMS-MES correspondence in the discrete-variable regime. The MMSs and MESs are main ingredients for the proof of existence of the additivity counterexample for the classical capacity of quantum channels~\cite{H09,FKM10,BH10}. Therefore, although there have been no practical suggestions for the proof to date, we can expect that Gaussian MMS-MES correspondence can be applied to the Gaussian channel-capacity problem. Since Gaussian states are well known and can be implemented in quantum optics, we consider this Gaussian MMS-MES correspondence as a tool for experimental proof of super-additivity of the classical channel-capacity problem. Here, we assume that a GMMS has the maximal von Neumann entropy in the same manner that a full-ranked $d$-dimensional MMS has the maximal entropy $\log d$.

\section{Gaussian MMS-MES correspondence via $g$-purifications} \label{g-pur}
We now briefly review the purification process in the discrete-variable case in order to set the stage for our investigation on the Gaussian CV case. Suppose that a mixed state $\rho_A$ can be decomposed by an orthonormal basis $\{\ket{i}_A\}_{i=1}^d$ such that $\rho_A=\sum_ip_i\proj{i}{i}_A$. To purify $\rho_A$, let us introduce an ancillary system $B$ with the orthonormal basis $\{\ket{i}_B\}_{i=1}^d$ whose dimension is same as that of the system $A$. If we define a pure state as
\begin{equation}
\ket{\psi}_{AB}:=\sum_{i=1}^d\sqrt{p_i}\ket{i}_A\ket{i}_B,
\end{equation}
then we naturally obtain the reduced density matrix of the system $A$ as ($\psi_{AB}:=\proj{\psi}{\psi}_{AB}$)
\begin{align}
\T_B(\psi_{AB})&=\sum_{i,j=1}^d\sqrt{p_ip_j}\proj{i}{j}_A\delta_{ij} \nonumber\\
&=\sum_ip_i\proj{i}{i}_A
=\rho_A.
\end{align}
Thus, for some fixed basis, $\ket{\psi}_{AB}$ is a purification of $\rho_A$. Now, suppose that $\ket{\Psi}_{AB}=\frac{1}{\sqrt{d}}\sum_{\ell=1}^d\ket{\ell}_A\otimes\ket{\ell}_B$ is a $d^2$-dimensional MES, we then obtain ($\Psi_{AB}:=\proj{\Psi}{\Psi}_{AB}$)
\begin{align}
\T_B(\Psi_{AB})&=\frac{1}{d}\sum_{\ell,m=1}^d\proj{\ell}{m}_A\delta_{\ell m} \nonumber\\
&=\frac{\1_A}{d}:=\rho_{d,\tn{MMS}}^A,
\end{align}
where $\1$ denotes the $d$-dimensional identity matrix. This implies that the MES $\ket{\Psi}_{AB}$ is one of the purifications of $\rho_{d,\tn{MMS}}^A$. The $d$-dimensional MMS $\rho_{d,\tn{MMS}}^A$ has an important property that is maximal von Neumann entropy, i.e.,  $S(\rho_{d,\tn{MMS}}^A)=-\T\left(\frac{\1_A}{d}\log\frac{\1_A}{d}\right)=\log d$, where $S(\varrho):=-\T\varrho\log\varrho$. This is crucial for quantum cryptographic protocols and the theory of quantum channel-capacity.

\begin{figure}
\centering
\includegraphics[width=1.01\columnwidth]{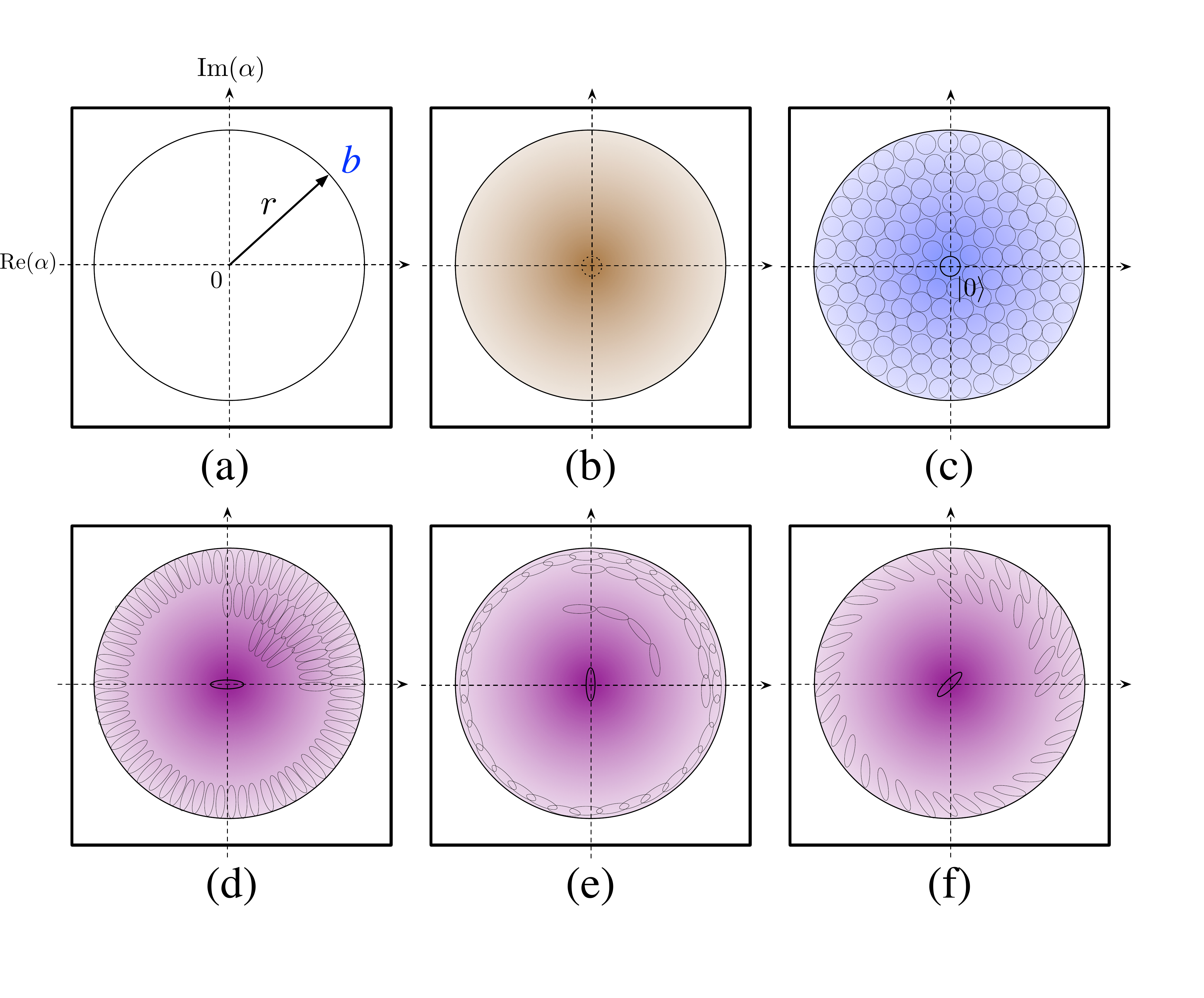}
\caption {%(color online). 
Several Gaussian maximally mixed state (GMMS) candidates in the phase space within a circle boundary $b$. Here all candidates are depicted in 2$d$ phase space, whose axes are Re($\alpha$) and Im($\alpha$). The radial component from the origin is expressed as $r$. Different colors represent the different kind of states and the density of color corresponds to the density of distribution function of state in the phase space. (a) ideal GMMS (uniform distribution), (b) thermal state with a given temperature, (c) Br\'{a}dler's continuous-variable MMS. Small circles are displaced coherent states, (d) squeezed GMMS with argument $\phi=0$ (e) $\phi=\frac{\pi}{2}$ and (f) $\phi=\frac{\pi}{4}$. Various shapes of squeezed circles are illustrated as direction of squeezing. These all exhibit different profiles within the boundary but become identical as the boundary tends to infinity. We note that the thermal state (b) has an infinite tail thus it tends to a uniform distribution only if the temperature approaches infinity.}
\label{Figure-1}
\end{figure}

We now consider the Gaussian CV case. In general, a $d$-mode Gaussian quantum system is described in $2d$-dimensional (real) symplectic phase space $\textnormal{Sp}(2d,\bb{R})$ and exists in the infinite dimensional Hilbert space with continuous eigenvalues of Gaussian observables~\cite{WPGCRSL12}. For convenience, we limit our discussion on the phase space with $d=1$, i.e., $\textnormal{Sp}(2,\bb{R})$. In the Gaussian regime, the concept of GMMS is not well-defined, in other words, the state cannot be uniquely specified. Prior to the main observation, we introduce an {\it ideal} GMMS, denoted as $\rho_{\tn{GMMS}}$ (See FIG.~\ref{Figure-1}(a)), which can be expressed by an equiprobable basis set, i.e., uniform distribution in the phase space. The distribution should also have a Gaussian profile, however, it becomes uniform only in the limiting case. All other candidate states should also tend to the uniform distribution as the boundary parameter approaches the limiting value. We must be aware that the state mentioned above is a quantum state, i.e., $\T(\rho_{\tn{GMMS}})=1$, but not an identity operator $\1$. For a bounded basis (parameters are not tending to infinity), $\1$ and MMS are identical up to a constant. However, it can be easily shown that $\T(\1)=\infty$ in the entire phase space because of its unbounded basis; we therefore need to consider a finite region of the phase space in which a circle of radius $b$ centered at origin and then take the limit to infinity. Note that FIG.~\ref{Figure-1} depicts several GMMS candidates in the phase space with some boundary $b$ from the origin. 

The firstly important candidate is thermal state, that can be written in the coherent state basis such as $\rho_{\tn{th}}(\bar{n})=\frac{1}{\bar{n}\pi}\int e^{-\frac{|\alpha|^2}{\bar{n}}}\proj{\alpha}{\alpha}d^2\alpha$, where $\bar{n}$ is the mean photon number and $\ket{\alpha}$ is a coherent state.
Unlike for all other cases, in the case of the thermal state (FIG.~\ref{Figure-1}(b)), the temperature (variance itself) is the regularizing parameter instead of a boundary of the phase space. Therefore, we can show that an infinite temperature (infinite variance) implies that the thermal state approaches the ideal MMS. If we introduce the Gaussian operations of displacement $\hat{D}(\alpha)=e^{\alpha\hat{a}^\dagger-\alpha^*\hat{a}}$ and squeezing $\hat{S}(\zeta)=e^{\frac{1}{2}(\zeta^*\hat{a}^2-\zeta\hat{a}^{\dagger2})}$ (where $\hat{a}$ and $\hat{a}^\dagger$ are the annihilation and the creation operators satisfying the commutation relation $[\hat{a},\hat{a}^\dagger]=1$), then a coherent and a squeezed coherent state, i.e., $\ket{\alpha}=\hat{D}(\alpha)\ket{0}\in\textnormal{Sp}(2,\bb{R})$ and $\ket{\alpha,\zeta}=\hat{S}(\zeta)\hat{D}(\alpha)\ket{0}\in\textnormal{Sp}(2,\bb{R})$, form an overcomplete set such that $\frac{1}{\pi}\int d^2\alpha\proj{\alpha}{\alpha}=\1$ and $\frac{1}{\pi}\int d^2\alpha\proj{\alpha,\zeta}{\alpha,\zeta}=\1$, respectively~\cite{BR97}. 

Moreover, it is important to note that the products of regularization of the convex combination of coherent or squeezed coherent states are GMMSs: for some (normalization) constants $k$ and $k'$, $\frac{1}{k}\sum_{i=1}^\infty \delta^2\alpha_i\proj{\alpha_i}{\alpha_i}=\rho_{\tn{GMMS}}^\alpha\in\textnormal{Sp}(2,\bb{R})$ and $\frac{1}{k'}\sum_{i=1}^\infty \delta^2\alpha_i\proj{\alpha_i,\zeta}{\alpha_i,\zeta}=\rho_{\tn{GMMS}}^{(\alpha,\zeta)}\in\textnormal{Sp}(2,\bb{R})$, respectively. For convenience, we omit the index $i$ and substitute the summation by the integral as $\delta^2\alpha_i\to 0$. Then, what we need to  investigate is whether $\rho_{\tn{GMMS}}=\rho_{\tn{GMMS}}^\alpha=\rho_{\tn{GMMS}}^{(\alpha,\zeta)}=\rho_{\textnormal{th}}\in\textnormal{Sp}(2,\bb{R})$.

Our main questions are: what is the purification of the ideal GMMS $\rho_{\tn{GMMS}}$ and is it a GMES? To answer these questions, we formulate a detailed Gaussian purification process (i.e., $g$-purification $\cl{P}_G$). (As shown in Ref.~\cite{FFLMP09}, it is known that there exists some class of perfect Gaussian multipartite entangled states.)

Our main result states that, for a given ideal GMMS $\rho_{\tn{GMMS}}^\gamma$ in a quantum system $A$ (with $\T(\rho_{\tn{GMMS}}^\gamma)=1$), there \emph{exist} GMESs obtained via $g$-purification processes such that
\begin{equation}
\cl{P}_G(\rho_{\tn{GMMS}}^\gamma)=\ket{\Gamma(\gamma)}_{AB},
\end{equation}
where $\gamma$ denotes a Gaussian parameter. Note that in general $\gamma$ depends on the parameters $n$ and $b$, i.e., $\gamma=\gamma(n,b)$, thus we should keep in mind that how this parameter $\gamma$ changes when we take the limit $n$ or $b$ to infinity.

To prove above statement, let us define a two-mode GMES that is the $g$-purified state in the Gaussian regime within the number basis as
\begin{equation}
\ket{\Gamma(\gamma)}_{AB}:=\sum_{n=0}^\infty \sqrt{f(\gamma)}\ket{n}_A\ket{n}_B,
\end{equation}
where the probability distribution function $f(\gamma)$ is related to the Gaussian parameter $\gamma$ such as displacement or squeezing. 
Note that $\lim_{n\to\infty}f(\gamma)$ converges to 0 ($\approx 1/{\infty}$), which implies that the distribution in the phase space is uniform (or flat). Thus, by using the partial trace on the ancillary system $B$, we obtain ($\Gamma(\gamma)_{AB}:=\proj{\Gamma(\gamma)}{\Gamma(\gamma)}_{AB}$)
\begin{align}
\T_B\big(\Gamma(\gamma)_{AB}\big)
&=\T_B\sum_{m,n=0}^\infty\sqrt{f(\gamma)f^*(\gamma)}\proj{nn}{mm}_{AB} \nonumber\\
&=\sum_{n=0}^\infty f(\gamma)\delta_{mn}\proj{n}{m}_A=\rho_{\tn{GMMS}}^\gamma,
\end{align}
where we use the fact that $f(\gamma)$ is a real-valued probability distribution. As mentioned above, the distribution of $f(\gamma)$ is uniform in the limit, and therefore, $\rho_{\tn{GMMS}}^\gamma$ corresponds to the GMMS with a Gaussian parameter $\gamma$. Finally, the $g$-purification of $\rho_{\tn{GMMS}}^\gamma$ in the system $A$ is a GMES $\ket{\Gamma(\gamma)}_{AB}$. This completes the proof.
\bigskip

It is important to note that if $\gamma=0$, then $\rho_{\tn{GMMS}}^0$ is the ideal GMMS, and if $\gamma$ is related to either displacement or squeezing operations, then $\rho_{\tn{GMMS}}^{(\alpha,\zeta)}$ is also a GMMS as an equiprobable convex sum of the coherent squeezed states. Although the GMMS is not unique, it exhibits the maximal entropy. Moreover, the orthonormal basis $\{\ket{n}_B\}_{n=0}^\infty$ is always related to another orthonormal set $\{\ket{n}_B'\}_{n=0}^\infty$ through the HJW theorem, so it is possible for a variety of GMESs to exist in the Gaussian regime. We now explicitly investigate three examples of the Gaussian MMS-MES correspondence.

\subsection{Thermal versus two-mode squeezed vacuum state} \label{thermal}
As a most basic Gaussian state, the first candidate of GMMS is the thermal state that maximizes the von Neumann entropy. The thermal state $\rho_{\tn{th}}$ is given by~\cite{BR97}
\begin{equation}
\rho_{\tn{th}}(\bar{n})
=\sum_{n=0}^\infty\frac{\bar{n}^n}{(\bar{n}+1)^{n+1}}\proj{n}{n},
\end{equation}
where $\bar{n}:=\T(\varrho\hat{a}^\dagger\hat{a})\ge0$ is the mean-photon number, and the state is also expressed by $\rho_{\tn{th}}(\bar{n})=\frac{1}{\bar{n}\pi}\int e^{-\frac{|\alpha|^2}{\bar{n}}}\proj{\alpha}{\alpha}d^2\alpha$ in the basis of coherent state previously.

Now, if we $g$-purify the thermal state, the resulting state is the well-known TMSV state. Suppose that the thermal state $\rho_{\tn{th}}$ is expressed in the number-state basis $\{\ket{n}_A\}_{n=0}^\infty$ such as $\rho_{\tn{th}}=\sum_{n}P_{\tn{th}}^n\proj{n}{n}_A$, where $P_{\tn{th}}^n:=\frac{(\bar{n})^{n}}{(\bar{n}+1)^{n+1}}$. To $g$-purify $\rho_{\tn{th}}$, let us introduce an ancillary system $B$ that has the same number basis as that of the system $A$ with  $\{\ket{n}_B\}_{n=0}^\infty$~\cite{Marian}. If we define a pure GMES as the TMSV state (via two-mode squeezing $\hat{S}_2(\zeta):=e^{\frac{\zeta}{2}(\hat{a}\hat{b}-\hat{a}^\dagger\hat{b}^\dagger)}$ in vacuum $\ket{0}_A\ket{0}_B$)
\begin{equation}
\ket{\Gamma(\zeta)}_{AB}=\sqrt{1-\lambda^2}\sum_{n=0}^\infty(-\lambda)^n\ket{n}_A\ket{n}_B,
\end{equation}
where $\lambda=\tanh\zeta\in[0,1]$, we obtain the reduced density matrix of $A$ as
\begin{align*}
\T_B\big(\Gamma(\zeta)_{AB}\big)
%&=\T_B\big[(1-\lambda^2)\sum_{n,m=0}^\infty(-\lambda)^{n+m}\proj{nn}{mm}_{AB} \big] \nonumber\\
%&=(1-\lambda^2)\sum_{n,m=0}^\infty(-\lambda)^{n+m}\delta_{nm}\proj{n}{m}_A \\
&=(1-\lambda^2)\sum_{n=0}^\infty(-\lambda)^{2n}\proj{n}{n}_A \nonumber\\
&=\sum_{n=0}^\infty P_{\tn{th}}^n\proj{n}{n}_A=\rho_{\tn{th}},
\end{align*}
where we let $\bar{n}=\sinh^2\zeta$.
This implies that the $g$-purification $\cl{P}_G(\rho_{\tn{th}})$ is GMES, i.e., TMSV state: $\cl{P}_G(\rho_{\tn{th}})=\ket{\Gamma(\zeta)}_{AB}$ with $f(\gamma):=(1-\lambda^2)(-\lambda)^{2n}$~\cite{Barnett}. It is known that the TMSV state has a Gaussian distribution in  the phase space~\cite{WPGCRSL12}. It is important to note that all above discussions of GMMSs and GMESs assume that the temperature (or average photon number) tends to infinity, whereas $f(\gamma)$ tends to zero (uniform distribution).

\subsection{Continuous-variable maximally mixed state and its $g$-purification} \label{CVMMS}
The concept of the continuous-variable maximally mixed state (CVMMS) was first introduced by Br\'{a}dler in 2005 for constructing a CV private quantum channel~\cite{B05,JKL15} as illustrated in FIG.~\ref{Figure-1}(c).

\begin{figure}
\centering
\includegraphics[width=1.01\columnwidth]{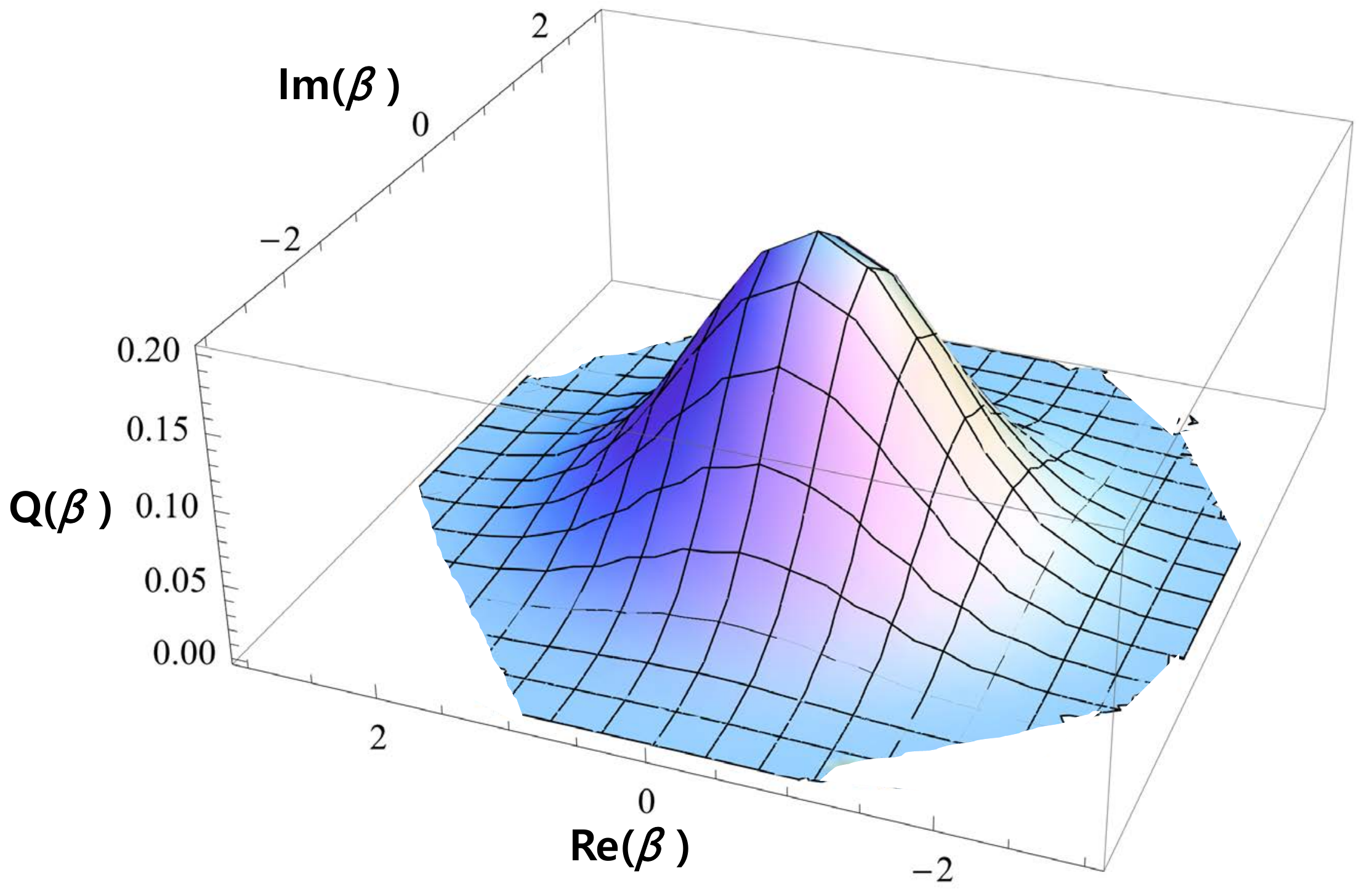}
\caption {%(color online). 
Husimi Q distribution of $\rho_{\tn{GMMS}}^{\alpha}|_b$ when $b=1$. This is simply the smoothed version of the Wigner distribution. As expected, a Gaussian profile in the phase space is observed. With increasing $b$, the function becomes closer to a uniform distribution, i.e., the equiprobable distribution corresponding to the maximally mixed state.}
\label{husimi}
\end{figure}

A CVMMS can be chosen as an integral performed over all possible single mode coherent state $\ket{\alpha}$ within the circle boundary of radius $r\le b$, imposing a physically motivated energy constraint. If $r>b$, the occurrence probability is 0. The coherent state can be expressed as $\ket{\alpha}=\hat{D}(\alpha)\ket{0}=e^{-|\alpha|^2/2}\sum_{n=0}^\infty\frac{\alpha^n}{\sqrt{n!}}\ket{n}$. We then obtain the GMMS~\cite{B05} using an equiprobable convex combination of coherent states up to normalization and describe it as
\begin{align} \label{eq:1b}
\rho_{\tn{GMMS}}^{\alpha}|_b&=\frac{1}{C}\int_b \proj{\alpha}{\alpha}d^2\alpha \nonumber\\
&=\frac{1}{b^2}\sum_{n=0}^\infty\left(1-\sum_{k=0}^n\frac{b^{2k}}{k!}e^{-b^2}\right)\proj{n}{n},
\end{align}
where the normalization constant is $C=\pi b^2$. We note that Br\'{a}dler's original paper denotes the GMMS as $\1_b$ to emphasize the boundary $b$ in the phase space. We now perform two investigations in order to confirm that $\rho_{\tn{GMMS}}^{\alpha}|_b$ is genuine a Gaussian quantum state:  
we examine whether $\rho_{\tn{GMMS}}^{\alpha}|_b$ has a unit trace and a Gaussian Wigner function. To check for the unit trace, we exchange the expression of the parenthesis in Eq.~(\ref{eq:1b}) into another form as
\begin{equation}
1-\sum_{k=0}^n\frac{b^{2k}}{k!}e^{-b^2}=\frac{\gamma (n+1,b^2)}{n!},
\end{equation}
where $\gamma (n+1,b^2)=\int^{b^2}_0 x^n e^{-x} dx$ is a lower incomplete Gamma function. Performing the infinite summation first and then doing the integral, we can easily show that $\T{\rho_{\tn{GMMS}}^{\alpha}|_b}=1$. To confirm the Gaussian character, it is better to consider the Husimi Q distribution $Q(\beta)=\frac{1}{\pi}\bra{\beta}\hat{\rho}\ket{\beta}$ instead of the Wigner function because the latter is not very smooth for finite $n$.  Husimi Q distribution is merely a Gaussian smoothing of the Wigner function $W(\alpha)$, i.e., $Q(\beta)=\frac{2}{\pi}\int d^2\alpha W(\alpha)e^{-2|\alpha-\beta|^2}$ and can be simply calculated in our case. It is numerically plotted in FIG.~\ref{husimi}, which clearly shows the Gaussian character of the distribution.

To purify the GMMS $\rho_{\tn{GMMS}}^{\alpha}|_b$, let us define a two-mode pure Gaussian state as
\begin{equation}
\ket{\Gamma(b)}_{AB}:=\sum_{n=0}^\infty \sqrt{f(b)}\ket{n}_A\ket{n}_B,
\end{equation}
where the distribution function $f(b)$ satisfies
\begin{equation}
f(b)=\frac{1}{b^2}\left(1-\sum_{k=0}^n\frac{b^{2k}}{k!}e^{-b^2}\right)\approx o\left(\frac{1}{b^2}\right),
\end{equation} 
and is uniform as long as $b\to\infty$.
If we assume that $n<\infty$, we obtain the GMMS
\begin{align*}
\T_B\big(\Gamma(r)_{AB}\big)
%&=\sum_{n=0}^\infty f(b)^2\delta_{nm}\proj{n}{m}_A \\
&=\frac{1}{b^2}\sum_{n=0}^\infty\left(1-\sum_{k=0}^n\frac{b^{2k}}{k!}e^{-b^2}\right)\proj{n}{n}_A \\
&=\rho_{\tn{GMMS}}^{\alpha}|_b.
\end{align*}
This result implies that the $g$-purification of the GMMS $\rho_{\tn{GMMS}}^{\alpha}|_b$ gives rise to a \emph{new} two-mode pure GMES in the Gaussian regime, i.e., $\cl{P}_G(\rho_{\tn{GMMS}}^{\alpha}|_b)=\ket{\Gamma(b)}_{AB}$. Since Eq.~(\ref{eq:1b}) is Gaussian, the $g$-purified state $\ket{\Gamma(b)}_{AB}$ is also Gaussian because the square-root of any Gaussian distribution generates another Gaussian distribution. Conversely, for some fixed number basis of the system $B$ as $\{\ket{n}_B\}_{i=0}^\infty$, the trace operation does not alter the Gaussian characteristics.

In Gaussian quantum information, the general Gaussian state is characterized by its first and second-order moments. Thus, it is natural to consider a squeezed GMMS as described below.

\subsection{Squeezed GMMS and its $g$-purification} \label{sGMMS}
Our final example is a squeezed GMMS and its $g$-purification process. We first introduce the squeezed GMMS as follows. In general, the squeezed coherent (SC) state is defined by~\cite{VW06} ($\zeta=se^{i\phi}$)
\begin{align}
\ket{\alpha,\zeta}
=&\hat{S}(\zeta)\hat{D}(\alpha)\ket{0} \nonumber\\
=&\sum_{n=0}^\infty\frac{(\nu/2\cosh s)^{n/2}}{\sqrt{\cosh s\cdot n!}}e^{-\frac{1}{2}\left(|\alpha|^2-\frac{\nu^*\alpha^2}{\cosh s}\right)} \nonumber\\
&\times H_n\left(\frac{\alpha}{\sqrt{2\nu\cosh s}}\right)\ket{n}, \label{eq:scstate}
\end{align} 
where $\nu=e^{i\phi}\sinh s$, and $\phi=\arg(\zeta)$ is the argument of squeezing parameter $\zeta$. In addition, note that $H_n(\cdot)$ represents the $n$th-degree of complex Hermite polynomials. For a given $\zeta$ (with some squeezing argument $\phi$), three cases of the squeezed GMMS are illustrated in FIG.~\ref{Figure-1}(d), (e), and (f). Exploiting Eq.~(\ref{eq:scstate}), we can define a squeezed GMMS with a boundary $r\le b$ defined by~\cite{JKL15}
\begin{align}
\rho_{\tn{GMMS}}^{(\alpha,\zeta)}|_b
&=\frac{1}{C}\int_b d^2\alpha\proj{\alpha,\zeta}{\alpha,\zeta} \nonumber\\
&=\frac{2\pi}{C}\sum_{m,n=0}^\infty\int_0^b\frac{(\tanh s/2)^{(m+n)/2}}{\cosh s\sqrt{m!n!}}e^{i\phi(m-n)/2} \nonumber\\
\times&e^{-Kr^2}H_m\left(\frac{re^{i(\theta-\frac{\phi}{2})}}{\sqrt{\sinh(2s)}}\right)H_n(\tn{c.c.})\proj{n}{m} \label{eq:sgmms1}\\
&=\frac{1}{b^2e^{Kb^2}}\sum_{n=0}^\infty\kappa_n(b,s,\phi)\proj{n}{n}, \label{eq:sgmms2}
\end{align}
where $C=\pi b^2$, $\theta$ is a relative angle between the SC states, $\tn{c.c.}$ denotes the complex conjugate, and $K:=1-\tanh s\cdot\cos(2\theta-\phi)$. In Eq.~(\ref{eq:sgmms1}), the relative angle $\theta$ converges to 0, because the angles between infinite SC states is quite small and approaches 0. Thus, $K=1-\tanh s\cdot\cos\phi$, and it depends only on the squeezing parameter $\zeta$. In Eq.~(\ref{eq:sgmms2}) (Eq.~(10) in Ref.~\cite{JKL15}), $\kappa_n(b,s,\phi)$ is an absolute constant with a small value (because for large $n$, $\kappa$ is proportional to a function of $\frac{\tn{exponential}}{\tn{factorial}}\ll1$), as obtained by integration over a delta function, the integration by parts, and the orthogonality condition of (complex) Hermite polynomials.
We note that the squeezing parameter $\zeta$ does not make the main contribution to the uniformity of the distribution. Therefore, we obtain the $g$-purification of $\rho_{\tn{GMMS}}^{(\alpha,\zeta)}|_b$ as described by
\begin{equation}
\cl{P}_G(\rho_{\tn{GMMS}}^{(\alpha,\zeta)}|_b)=\ket{\Gamma(b,\zeta)}_{AB},
\end{equation}
where $\ket{\Gamma(b,\zeta)}_{AB}:=\sum_{n=0}^\infty\sqrt{f(b,\zeta)}\ket{n}_A\ket{n}_{B}$, which is another new GMES with the uniform phase space distribution $f(b,\zeta)=\frac{1}{b^2e^{Kb^2}}\kappa_n(b,s,\phi)\approx o\left(\frac{1}{b^2e^{Kb^2}}\right)$. We can also observe that $f(b,\zeta)$ in the squeezed GMMS has an almost uniform (or flat) distribution, which is more uniform as $b\to\infty$ than the distribution of $f(b)$ in CVMMS owing to the exponential term. Since displacement and squeezing are Gaussian operations, our squeezed GMMS is also a Gaussian state by the definition of Gaussian operation. Consequently, its purified state is also Gaussian; hence, we can conclude that any purification of a Gaussian state will always give rise to a Gaussian state.

Furthermore, we conjecture that the distance between the purified states of squeezed GMMS and CVMMS (i.e., GMES and CVMES) is very small. This arises from the fact that for sufficiently small and positive $\eps$, the distance between the squeezed GMMS $\rho_{\tn{GMMS}}^{(\alpha,\zeta)}|_b$ and the Br\'{a}dler's CVMMS $\rho_{\tn{GMMS}}^\alpha|_b$ is always sufficiently small as shown in Ref.~\cite{JKL15} by unitary invariance and the norm convexity: $\|\rho_{\tn{GMMS}}^{(\alpha,\zeta)}|_b-\rho_{\tn{GMMS}}^\alpha|_b\|_2\le\eps$. Note that $\|\cdot\|_2$ denotes the Hilbert-Schmidt norm such that $\|M\|_2=\sqrt{M^\dagger M}$ for any matrix $M$. When we choose $b\to\infty$, we also expect that the distributions of squeezed GMMS and CVMMS are approaching to that of the ideal GMMS.

\section{Conclusions}\label{conclusion}
In this work, we address the basic question of what the Gaussian MMS and MES are. Despite its simplicity, heretofore, this question has not been clearly answered. We present several GMMS candidates, and propose a new method to obtain GMESs by using the concept of Gaussian MMS-MES correspondence. For a given GMMS, we can always construct the GMES via the $g$-purification; thus, this approach is a very simple yet powerful tool for CV quantum information processing. This procedure may shed light on the subject of theoretical Gaussian quantum information processing as well as plausible experimental realization with current technologies in the sense that Gaussian states and Gaussian operations are much easier to implementable and manipulable than general one~\cite{WPGCRSL12,Wang2007}.
We present $g$-purifications for several Gaussian MMSs, including the well-known TMSV state and the thermal state, and derive two new candidates of Gaussian MESs from Br\'{a}dler CVMMS and the squeezed GMMS. Furthermore, we observe that any $g$-purified states of GMMSs also give rise to Gaussian states. Thus, it is possible to state that a Gaussian purification process preserves the Gaussian characteristics of a state.

Several open questions still remain. For example, this work did not address the origin of the non-uniqueness of the Gaussian MMS. It turns out that the answer to this rather subtle problem may lead to a breakthrough for the understanding of ubiquitous singularities such as infinite energy and the infinite squeezing for the maximal entanglement.

Finally, the possible experimental testing of the degree of additivity violation of classical capacity in quantum channel remains to be explored. The actual observation of additivity violation is very important for efficient quantum communication, but there have been no significant (experimental) result addressing this issue to date although there is an experimental suggestion in the case of multiple-access Gaussian channel~\cite{MAC}. In addition to the discrete version for the classical capacity counterexample in quantum channels~\cite{H09}, we expect that a similar counterexample exists in the CV Gaussian regime. As mentioned above, there are significant advantages in the CV Gaussian regime over the discrete-variable case for the feasible experiments. Clearly the relation between the Gaussian MMS and Gaussian MES will enable a breakthrough for the experiments addressing additivity violation in the quantum channels.

\section*{Acknowledgements}
The authors thank Su-Yong Lee, Jaewan Kim, In Young Lim, Dong Pyo Chi, and Hyunseok Jeong for their valuable comments and support. The authors acknowledge financial support by the National Research Foundation of Korea (NRF) grant funded by the Korea government (MSIP) (Grant No. 2010-0018295). K.J. acknowledges the Associate Member Program funded by the Korea Institute for Advanced Study (KIAS).

\end{document}